\documentclass[conference]{IEEEtran}
\usepackage{amsmath,amssymb,bm,cite,algorithm,algpseudocode,amsthm}
\usepackage{graphicx, xcolor}
\usepackage{caption}

\newcommand{\h}{\mathbf{h}}

\newcommand{\n}{\mathbf{n}}

\newcommand{\s}{\mathbf{s}}

\renewcommand{\u}{\mathbf{u}}

\newcommand{\y}{\mathbf{y}}

\newcommand{\0}{\mathbf{0}}

\newcommand{\A}{\mathbf{A}}

\newcommand{\C}{\mathbf{C}}
\newcommand{\D}{\mathbf{D}}

\newcommand{\F}{\mathbf{F}}
\newcommand{\G}{\mathbf{G}}
\renewcommand{\H}{\mathbf{H}}
\newcommand{\I}{\mathbf{I}}

\renewcommand{\L}{\mathbf{L}}

\newcommand{\U}{\mathbf{U}}
\newcommand{\V}{\mathbf{V}}

\newcommand{\Real}{\mbox{$\mathbb{R}$}}
\newcommand{\Compl}{\mbox{$\mathbb{C}$}}

\newcommand{\tr}{\mathrm{tr}}

\begin{document}
\title{Joint Design of Multi-Tap Analog Cancellation and Digital Beamforming for Reduced Complexity \\Full Duplex MIMO Systems \vspace{-5mm}}
\author{
\IEEEauthorblockN{George~C.~Alexandropoulos and Melissa Duarte}
\IEEEauthorblockA{Mathematical and Algorithmic Sciences Lab, Huawei Technologies France SASU, 92100 Boulogne-Billancourt, France}
emails: \{george.alexandropoulos, melissa.duarte\}@huawei.com}

\maketitle

\begin{abstract}
	Incorporating full duplex operation in Multiple Input Multiple Output (MIMO) systems provides the potential of boosting throughput performance. However, the hardware complexity of the analog self-interference canceller scales with the number of transmit and receive antennas, thus exploiting the benefits of analog cancellation becomes impractical for full duplex MIMO transceivers. In this paper, we present a novel architecture for the analog canceller comprising of reduced number of taps (tap refers to a line of fixed delay and variable phase shifter and attenuator) and simple multiplexers for efficient signal routing among the transmit and receive radio frequency chains. In contrast to the available analog cancellation architectures, the values for each tap and the configuration of the multiplexers are jointly designed with the digital beamforming filters according to certain performance objectives. Focusing on a narrowband flat fading channel model as an example, we present a general optimization framework for the joint design of analog cancellation and digital beamforming. We also detail a particular optimization objective together with its derived solution for the latter architectural components. Representative computer simulation results demonstrate the superiority of the proposed low complexity full duplex MIMO system over lately available ones.


\end{abstract}

\begin{IEEEkeywords}
Analog cancellation, beamforming, combining, full duplex, MIMO, multi-user systems, optimization, precoding.
\end{IEEEkeywords}

\section{Introduction}\label{sec:Introduction}
In band full duplex, also known shortly as Full Duplex (FD), is a candidate technology for fifth generation (5G) wireless systems because of the potential spectral efficiency gains that can be achieved through simultaneous uplink and downlink communication within the entire frequency band \cite{Sab14_all, Am45G14}. An FD radio can transmit and receive at the same time and same frequency resource unit, consequently, it can double the spectral efficiency achieved by a half duplex radio. Current wireless systems exploit Multiple Input Multiple Output (MIMO) communication, where increasing the number of transmit and receive antennas can increase the spatial Degrees of Freedom (DoF), hence boosting spectral efficiency. Combining FD with MIMO communication can provide further spectral efficiency gains \cite{Rii11_all, Nguyen2013_all, Bha14, SofNull_2016, Tam2016_all, GA2016_all}. Thus, enabling FD MIMO technology, for small to large antenna array systems, is of high interest in order to achieve the demanding throughput requirements of 5G wireless communication systems. 

An FD radio suffers from self interference, which is the signal transmitted by the FD radio Transmitter (TX) that leaks to the FD radio Receiver (RX). At the RX of the FD radio, the power of the self-interference signal can be many times stronger than the power of the received signal of interest (which is transmitted from another radio). Consequently, self interference can severely degrade the reception of the signal of interest, and thus self-interference mitigation is required in order to maximize the spectral efficiency gain of the FD operation. As the number of antennas increases, mitigating self interference becomes more challenging, since more antennas naturally result in more self-interference components. For the case of a Single Input Single Output (SISO) FD node, it has been demonstrated \cite{Bha13_all,Dua14_all} that significant self-interference mitigation can be achieved via a combination of analog and digital cancellation techniques, where an estimate of the received self interference is subtracted from the received signal (which is the sum of the self-interference signal and the signal of interest). A straightforward extension of self-interference mitigation solutions used in SISO FD to the case of MIMO FD can be envisioned. However, the hardware resources required for analog self-interference cancellation become a main bottleneck, since they scale with the number of antenna elements. Thus, recent works have proposed only digital self-interference mitigation for FD MIMO \cite{Rii11_all,SofNull_2016}. These approaches exploit the availability of multiple antennas at the FD node in order to provide self-interference mitigation via digital beamforming; such an approach is known as spatial suppression. However, as has been pointed out, spatial suppression approaches often result in lower rates for both the outgoing and incoming signals of interest, since some of the available spatial DoF are solely devoted for mitigating self interference.

In this paper, we propose a novel architecture for analog self-interference cancellation and a novel optimization framework for joint design of the analog canceller and TX and RX digital beamforming parameters. The new architecture for analog cancellation consists of multi-tap analog canceller hardware, where the number of taps does not increase with the number of TX or RX antenna elements. The number of taps can be chosen offline as a function of size constraints, cost per tap, or other constraints on the analog canceller hardware. This simplified analog canceller architecture is enabled via the use of multiplexers, which allow flexible connectivity between the taps and the TX and RX antennas. The settings of taps and the configurations of multiplexers is computed via our proposed optimization framework. The flexible signal routing via multiplexers enables the use of reduced taps in an optimized way, since the taps will be used between the subset of TX and RX antennas where they are mostly beneficial. The digital beamformer and analog canceller parameters are thus designed by taking into account each others capabilities, hence the burden of self-interference mitigation is split between digital beamforming and analog cancellation. We note that the related work \cite{Rii11_all} has considered joint design of digital beamforming and analog cancellation, however these and related solutions \cite{Atz16_all,Zha12_all} assume underlying analog canceller hardware as in \cite{Bha13_all, Bha14}, which scales with the number of transmit and receive antennas. As our simulation results will show, the proposed analog canceller architecture together with our novel co-design of analog cancellation and TX and RX digital beamforming is capable of achieving higher rates with less hardware compared to the State-of-the-Art (SotA) FD MIMO solutions. 

\textit{Notation:} Vectors and matrices are denoted by boldface lowercase and boldface capital letters, respectively. The transpose and Hermitian transpose of $\mathbf{A}$ are denoted by $\mathbf{A}^{\rm T}$ and $\mathbf{A}^{\rm H}$, respectively, and $\det(\mathbf{A})$ is the determinant of $\mathbf{A}$, while $\mathbf{I}_{n}$ ($n\geq2$) is the $n\times n$ identity matrix and $\mathbf{0}_{n}$ ($n\geq2$) is the $n$-element zero vector. $\|\mathbf{a}\|$ stands for the Euclidean norm of $\mathbf{a}$ and ${\rm diag}\{\mathbf{a}\}$ denotes a square diagonal matrix with $\mathbf{a}$'s elements in its main diagonal. $[\mathbf{A}]_{i,j}$, $[\mathbf{A}]_{(i,:)}$ and $[\mathbf{A}]_{(:,j)}$ represent $\mathbf{A}$'s $(i,j)$-th element, $i$-th row, and $j$-th column, respectively, while $[\mathbf{a}]_{i}$ denotes the $i$-th element of $\mathbf{a}$. $\Real$ and $\Compl$ represent the real and complex number sets, respectively, $\mathbb{E}\{\cdot\}$ is the expectation operator, and $|\cdot|$ is the amplitude of a complex number.


\section{Proposed FD MIMO Framework}\label{sec:FD_MIMO}
\begin{figure*}
	\begin{center}
	\includegraphics[width=7in]{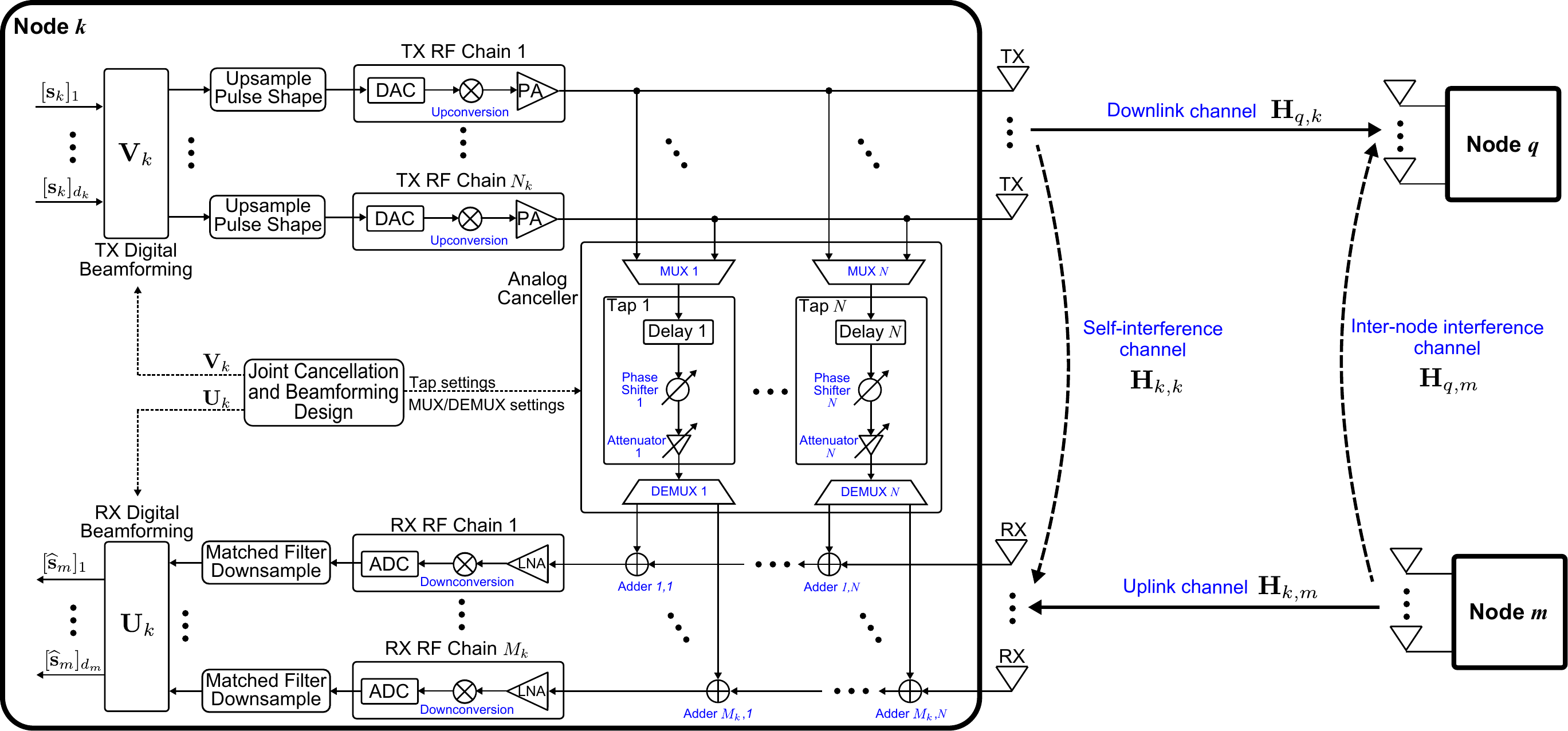}
	\caption{The FD MIMO node $k$ communicates with the two half duplex multi-antenna nodes $q$ and $m$, the former in the downlink and the latter in the uplink communication. The figure shows in detail the hardware components of the proposed analog canceller architecture at node $k$. The analog canceller consists of $N$ taps, which are connected via multiplexers (MUXs)/demultiplexers (DEMUXs) to the outputs of the TX radio frequency (RF) chains and the inputs of the RX RF chains. With the term ``tap'' we denote a line of fixed delay and variable phase shifter and attenuator. The figure also depicts the processing blocks dedicated to TX digital beamforming, RX digital beamformning, and to the design of the joint analog cancellation and digital beamforming.}
	\label{fig:FD_MIMO}
	\end{center}
\end{figure*}

The proposed FD MIMO framework including a novel architecture for the analog self-interference canceller and a joint design of its components together with the TX and RX digital beamforming filters is detailed in the following. We consider a communication system where a FD MIMO node $k$ communicates concurrently with a multi-antenna node $q$ in the downlink and a multi-antenna node $m$ in the uplink, as shown in Fig$.$~\ref{fig:FD_MIMO}. Since we intend at investigating FD operation at a single node, we henceforth assume that nodes $q$ and $m$ operate in half duplex mode. The novel hardware features of our analog canceller are described in Sec$.$~\ref{subsec:Analog_Canceller}, whereas Sec$.$~\ref{subsec:Signal_Model} presents the considered signal model that is later used on Sec$.$~\ref{subsec:BDC} for introducing the proposed general optimization framework for the co-design of analog cancellation and digital beamforming. 

\subsection{Novel Analog Canceller Architecture}\label{subsec:Analog_Canceller} 


According to our novel analog canceller architecture illustrated in Fig.~\ref{fig:FD_MIMO} for the FD MIMO node $k$, $N$ canceller taps are applied, via multiplexers (MUXs) and demultiplexers (DEMUXs), between the $N_k$ TX Radio Frequency (RF) chains and the $M_k$ RX RF chains, respectively. One way of implementing analog RF MUX/DEMUXs is through RF switches. With the term `tap' we denote a a fixed delay-variable phase shifter-variable attenuator line, as considered in \cite{Kol16_all}. A TX RF chain consists of a Digital to Analog Converter (DAC), a mixer which upconverts the signal from baseband to RF, and a Power Amplifier (PA). An RX RF chain consists of a Low Noise Amplifier (LNA), a mixer which downconverts the signal from RF to baseband, and an Analog to Digital Converter (ADC). At the TX side, upsample and pulse shape processing are used to prepare the baseband signal for DAC sampling and RF transmission. At the RX side, a corresponding matched filter and downsampling is performed. The details of the TX and RX digital beamforming blocks will be explained in later sections.    

We now focus on the description of the proposed analog canceller architecture. As shown in Fig.~\ref{fig:FD_MIMO}, the input of each analog canceller tap is connected to a corresponding $N_k$-to-1 MUX which allows routing of any of the $N_k$ TX RF chain signals to the input of the tap. The connection from each TX RF chain to each MUX input can be done via power dividers or directional couplers \cite{Kol16_all}. The signal that inputs to a tap undergoes a delay, phase shift, and attenuation, and this generates as an output an analog cancelling signal. The output of each tap is connected to an 1-to-$M_k$ DEMUX, which routes the cancelling signal at the output of the tap to one of the adders located just before the RX RF chains. There are a total of $M_kN$ such adders and we use ``Adder $i,j$'' to label the adder that connects DEMUX $j$ to RX RF chain $i$. Thus, the signal input to the $i$-th RX RF chain is the result of adding $N$ cancelling signals to the signal received at the $i$-th RX antenna element. Since the adders are connected to DEMUXs, some of the adders may have zero in one of the inputs depending on the DEMUXs' settings. The adders before the RX RF chains can be implemented via power combiners or directional couplers. 



The use of MUXs/DEMUXs for signal routing is a novel feature of our analog self-interference canceller. Moreover, our proposed co-design of analog cancellation and digital beamforming, which will be explained in Section~\ref{subsec:BDC}, jointly optimizes the TX and RX digital beamformers, the MUX/DEMUX configurations and the tap settings (i$.$e$.$, values for the phase shifters and attenuators). This joint design splits the burden of self-interference cancellation between the digital beamformers and the analog canceller, thus allowing the use of a reduced number of taps for the analog canceller, compared to the number of taps required by the designs in \cite{Bha13_all,Bha14,Kol16_all}, which require at least one tap between each TX RF chain and each RX RF chain. For our proposed analog canceller design, the total number of taps $N$ is flexible and can be chosen offline as a function of size constrains, cost per tap, or other constraints on the analog canceller hardware. The TX and RX digital beamformers and analog canceller will adapt to each others capabilities via their joint design.

\subsection{Signal Model}\label{subsec:Signal_Model} 
Suppose that the FD MIMO node $k$ in Fig$.$~\ref{fig:FD_MIMO} is equipped with $N_k$ TX antenna elements and $M_k$ RX antenna elements, each attached to a dedicated RF chain as depicted in the figure. The half duplex multi-antenna nodes $q$ and $m$ are assumed to have $M_q$ and $N_m$ antennas, respectively, with each antenna connected to an RF chain. For presentation clarity purposes, we assume narrowband flat fading channels for our signal model, and extensions for wideband frequency selective channels are currently part of our future work. All nodes are considered capable of performing digital beamforming, which for simplicity we assume hereinafter to be realized with linear filters. In particular, we assume that node $k$ makes use of the precoding matrix $\V_k\in\Compl^{N_k\times d_k}$ for processing its unit power symbol vector $\s_k\in\Compl^{d_k\times1}$ (chosen from a discrete modulation set) before transmission. The dimension of $\s_k$ satisfies $d_k\leq\min\{M_q,N_k\}$, which complies with the available spatial DoF for the downlink $M_q\times N_k$ MIMO channel. Similarly, node $m$ processes its unit power symbol vector $\s_m\in\Compl^{d_m\times1}$ (chosen again from a discrete modulation set) with a precoding matrix $\V_m\in\mathcal{C}^{N_m\times d_m}$, where $d_m\leq\min\{M_k,N_m\}$. Both the downlink and uplink transmissions are power limited according to $\mathbb{E}\{\|\V_k\s_k\|^2\}\leq {\rm P}_k$ and $\mathbb{E}\{\|\V_m\s_m\|^2\}\leq {\rm P}_m$. 

Upon signal reception at the FD MIMO node $k$, analog self-interference cancellation is first applied to the signals received at the RX antenna elements before these signals enter to the RX RF chains, as shown in Fig$.$~\ref{fig:FD_MIMO}. We utilize the notation $\C_k\in\Compl^{M_k\times N_k}$ to represent the analog processing realized by the analog canceller; $\C_k$ captures the configuration of the MUXs/DEMUXs and the canceller tap values. More specifically, we model $\C_k$ in baseband representation as the following cascade of three matrices 
\begin{equation}\label{Eq:C_k}
\C_{k} \triangleq \L_3\L_2\L_1,
\end{equation}
where $\L_1\in\Real^{N \times N_k}$, $\L_2\in \Compl^{N\times N}$, and $\L_3\in\Real^{M_k\times N}$. The elements $[\L_1]_{i,j}$ with $i=1,2,\ldots,N$ and $j=1,2,\ldots,N_k$, and $[\L_3]_{i,j}$ with $i=1,2,\ldots,M_k$ and $j=1,2,\ldots,N$ take the binary values $0$ or $1$, and it must hold that 
\begin{subequations}\label{Eq:L_1_L_3}
\begin{equation}\label{Eq:L_1}
\sum_{j=1}^{N_k}[\L_1]_{i,j} = 1 \,\,\forall i=1,2,\ldots,N,
\end{equation}
\begin{equation}\label{Eq:L_3}
\sum_{i=1}^{M_k}[\L_3]_{i,j} = 1 \,\,\forall j=1,2,\ldots,N.
\end{equation}
\end{subequations}
The $i$-th row of $\L_1$ indicates the MUX configuration at the input of the $i$-th tap of the canceller, while the $i$-th column of $\L_3$ shows the DEMUX configuration at the output of the $i$-th tap of the canceller. The $\L_2$ in \eqref{Eq:C_k} is a diagonal matrix whose complex entries represent the attenuation and phase shift of the canceller taps; the magnitude and phase of the element $[\L_2]_{i,i}$ with $i=1,2,\ldots,N$ specify the attenuation and phase of the $i$-th tap. Recall that the tap delays in each canceller tap are fixed and since we focus on a narrowband system, we model the effects of the $i$-th tap delay as a phase shift that is incorporated to the phase of $[\L_2]_{i,i}$. Following the above definitions, the baseband received signal $\y_{q}\in\Compl^{M_q\times 1}$ at node $q$ can be mathematically expressed as
\begin{equation}\label{Eq:Received_q}
\y_{q} \triangleq \H_{q,k}\V_k\s_{k} + \H_{q,m}\V_{m}\s_{m} + \n_{q}, 
\end{equation}
where $\H_{q,k}\in\Compl^{M_q\times N_k}$ is the downlink channel gain matrix (i$.$e$.$, between the nodes $q$ and $k$), $\H_{q,m}\in\Compl^{M_q\times N_m}$ denotes the channel gain matrix for inter-node interference (i$.$e$.$, between nodes $q$ and $m$), and $\n_{q}\in\Compl^{M_q\times 1}$ represents the additive white Gaussian noise (AWGN) at node $q$ with variance $\sigma_{q}^{2}$. By assuming that the digitally converted downsampled output signals of the RX RF chains at node $k$ are linearly processed in baseband by the combining matrix $\U_k\in\Compl^{d_m\times M_k}$, the estimated symbol vector $\hat{\s}_m\in\Compl^{d_m\times 1}$ for $\s_m$ is derived as  
\begin{equation}\label{Eq:Estimated_m} 
\hat{\s}_m \triangleq  \U_k\left(\y_k+\overline{\y}_k + \widetilde{{\y}}_k+\n_{k}\right), 
\end{equation}
where the complex-valued $M_k$-element vectors $\y_k$, $\overline{\y}_k$, and $\widetilde{{\y}}_k$ are the baseband representations of the received signal of interest, received self-interference signal, and output signal of the analog canceller, respectively, at node $k$. In addition, $\n_{k}\in\Compl^{M_k\times 1}$ denotes the received AWGN at node $k$ with variance $\sigma_{k}^{2}$. The vector $\y_k$ in \eqref{Eq:Estimated_m} is given by  
\begin{equation}\label{Eq:y_k_SoI}
\y_{k} \triangleq \H_{k,m}\V_{m}\s_{m},
\end{equation}
where $\H_{k,m}\in\Compl^{M_k\times N_m}$ is the uplink channel gain matrix (i$.$e$.$, between the nodes $k$ and $m$), while $\overline{\y}_k$ is obtained as 
\begin{equation}\label{Eq:y_k_SI}
\overline{\y}_{k} \triangleq \H_{k,k}\V_{k}\s_{k},
\end{equation}
with $\H_{k,k}\in\Compl^{M_k\times N_k}$ denoting the self-interference channel seen at the RX antennas of node $k$ due to its own downlink transmission. Finally, $\widetilde{{\y}}_k$ is given by   
\begin{equation}\label{Eq:y_k_AC}
\widetilde{\y}_{k} \triangleq \C_k\V_k\s_{k}.
\end{equation}

\subsection{Novel Joint Design of Analog Cancellation and Digital Beamforming}\label{subsec:BDC}  
A core element of the proposed FD MIMO framework is the joint design of the components of our analog canceller architecture together with the TX and RX digital beamforming blocks in order to satisfy certain performance objectives. Let us focus on the signal model presented in Sec$.$~\ref{subsec:Signal_Model} and on the co-design of $\C_k$, $\V_k$, and $\U_k$.  We define the scalar performance function $f$ having as inputs the analog canceller matrix $\C_k$, the digital precoding matrix $\V_k$, and the digital combining matrix $\U_k$. Note that $f$ may represent a sole performance objective, such as the throughput performance of the FD MIMO operation, or be a multi-objective function. Our general optimization framework for the joint design of $\C_k$, $\V_k$, and $\U_k$ can be mathematically expressed by the following general optimization problem:
\begin{equation*}\label{eq:optim}
\begin{split}
  \mathcal{OP}: &\max_{\C_k,\V_k,\U_k} f\left(\C_k,\V_k,\U_k\right)
	\\& \hspace{0.6cm}\textrm{s.t.}~~\tr\{\V_k\V_k^{\rm H}\}\leq{\rm P}_k, \hspace{2.8cm}({\rm C1})
	\\& \hspace{1.26cm}               \C_k = \L_3\L_2\L_1 \,\,{\rm with}\,\,\eqref{Eq:L_1},\eqref{Eq:L_3},{\rm and}
	\\& \hspace{2.095cm}              [\L_2]_{i,j}=0\,\,{\rm for}\,\,i\neq j, \hspace{1.5cm}({\rm C2}) 
	\\& \hspace{1.26cm}               \|(\H_{k,k}+\C_k)\V_k\s_k\|^2 \leq \lambda_{\rm A}, \hspace{1.2cm}({\rm C3})
	\\& \hspace{1.26cm}               \|\U_k(\H_{k,k}+\C_k)\V_k\s_k\|^2 \leq \lambda_{\rm D}, \hspace{0.7cm}({\rm C4})
\end{split}
\end{equation*}
where constraint $({\rm C1})$ relates to the average transmit power at node $k$ and constraint $({\rm C2})$ refers to the hardware capabilities of the analog canceller. Moreover, constraint $({\rm C3})$ imposes the threshold $\lambda_{\rm A}\in\Real$ on the instantaneous residual self interference after analog cancellation, while constraint $({\rm C4})$ sets the threshold $\lambda_{\rm D}\in\Real$ on the instantaneous residual self interference after applying analog cancellation and RX digital beamforming. 

The main novel components of the proposed joint analog cancellation and digital beamforming design described in $\mathcal{OP}$ are summarized as follows. First, the digital beamforming design takes explicit account of the available number of taps $N$ of the analog canceller. Although some available beamforming solutions \cite{Rii11_all,Atz16_all} for FD MIMO systems take into account the presence of an analog self-interference canceller, the details of its hardware limitations are excluded from the beamforming design. Second, the proposed FD MIMO framework is the only one that considers the case where $N<\min\{M_k,N_k\}$, i$.$e$.$, the available number of analog canceller taps may be smaller than both the numbers of TX and RX RF chains. This is an important feature for FD MIMO deployments, since current analog canceller solutions \cite{Bha14} require very large numbers of taps of the order of $M_kN_k$. Our framework has the advantage of a more optimized utilization of the spatial DoF offered by the multiple antennas. For example, if the analog canceller consists of only $N=1$ tap, then its cancellation capabilities are very limited, and more spatial DoF need to be devoted from the TX and RX beamforming blocks for meeting the thresholds $\lambda_{\rm A}$ and $\lambda_{\rm D}$ in $({\rm C3})$ and $({\rm C4})$. On the other extreme, if $N$ can be afforded to be large, the digital beamforming design may exploit the fact that a significant part of self-interference mitigation is handled by the analog canceller, and make use of more spatial DoF for improving the quality of the incoming and outgoing signals of interest. 

\section{An Example FD MIMO Design}\label{sec:Solution}
Capitalizing on the general optimization framework for the co-design of $\C_k$, $\V_k$, and $\U_k$ described in Sec$.$\ref{subsec:BDC}, we present in this section an example joint design of analog cancellation and digital beamforming. We consider the signal model presented in Sec$.$\ref{subsec:Signal_Model} for the special case where the half duplex single antenna node $m$ transmits one symbol stream without precoding in the uplink; particularly, we set $N_m=d_m=1$ and $\V_m={\rm P}_m^{1/2}$. We also introduce the notation $s_m\in\Compl$ for the sole unit power symbol stream per transmission of node $m$, and represent by $\h_{k,m}\in\Compl^{M_k\times 1}$ the channel gain matrix between nodes $k$ and $m$. Finally, we assume that there is no inter-node interference between nodes $q$ and $m$ due to, for example, appropriate node scheduling \cite{GA2016_all} for the FD operation of node $k$. This translates to setting the channel gain matrix $\h_{q,m}\in\Compl^{M_q\times 1}$ between these two nodes as $\h_{q,m}=\0_{M_q}$. Using the latter definitions, expression \eqref{Eq:Estimated_m} that describes the estimation for $s_m$ can be rewritten as    
\begin{equation}\label{Eq:Estimated_m1} 
\hat{s}_m \triangleq  \u_k\left({\rm P}_m^{1/2}\h_{k,m}s_m + \widetilde{\H}_{k,k}\V_k\s_k + \n_{k}\right), 
\end{equation} 
where $\u_k\in\Compl^{1\times M_k}$ represents the combining vector at the FD node $k$ and $\widetilde{\H}_{k,k}\in\Compl^{M_k\times N_k} $ denotes the effective self-interference channel after performing analog cancellation, which is defined as $\widetilde{\H}_{k,k}\triangleq\H_{k,k}+ \C_k$. The average post-processing signal-to-interference-plus-noise ratio (SINR) of $\hat{s}_m$, i$.$e$.$, after applying the RX digital beamformer at node $k$ and averaging over all transmitted symbols $s_m$, is given as a function of $\C_k$, $\V_k$, and $\u_k$ by  
\begin{equation}\label{Eq:SINR}
\gamma_k\left(\C_k,\V_k,\u_k\right) = \frac{{\rm P}_m|\u_k\h_{k,m}|^2}{\|\u_k\widetilde{\H}_{k,k}\V_k\|^2+\sigma_k^2\|\u_{k}\|^2}.
\end{equation}
  
An important performance objective function $f$ for the considered system is the FD data rate defined as the sum rate of the downlink and uplink communications. We therefore focus on designing $\C_k$, $\V_k$, and $\u_k$ via the solution of the following optimization problem:
\begin{equation*}\label{eq:optim1}
\begin{split}
  \mathcal{OP}1: &\max_{\C_k,\V_k,\u_k} \mathcal{R}_{\rm DL}\left(\V_k\right) + \mathcal{R}_{\rm UL}\left(\C_k,\V_k,\u_k\right)
	\\& \hspace{0.57cm}\textrm{s.t.}~~({\rm C1}),\,\,({\rm C2}),\,\,\|\u_k\|^2=1,
	\\& \hspace{1.3cm}              \|[\widetilde{\H}_{k,k}\V_k]_{(i,:)}\|^2\leq\lambda_{\rm A} \hspace{0.15cm} \forall i=1,2,\ldots,M_k.
\end{split}
\end{equation*}
In the latter problem, the achievable downlink rate $\mathcal{R}_{\rm DL}(\cdot)$ is a function of only $\V_k$ and is given by  
\begin{equation}\label{Eq:DL_Rate}
\mathcal{R}_{\rm DL}\left(\V_k\right) = \log_2\left(\det\left(\I_{M_q}+\sigma_q^{-2}\H_{q,k}\V_k\V_k^{\rm H}\H_{q,k}^{\rm H}\right)\right),
\end{equation} 
whereas the uplink rate $\mathcal{R}_{\rm UL}(\cdot)$ is a function of $\C_k$, $\V_k$, and $\u_k$, and is derived as
\begin{equation}\label{Eq:UL_Rate}
\mathcal{R}_{\rm UL}\left(\C_k,\V_k,\u_k\right) = \log_2\left(1+\gamma_k\left(\C_k,\V_k,\u_k\right)\right).
\end{equation}

Note that in the formulation of $\mathcal{OP}1$ we have relaxed the constraint $({\rm C3})$ concerning the instantaneous residual self interference after analog cancellation that appears in the general $\mathcal{OP}$ to an average power per RX RF chain constraint, where the average is taken over all possible transmit symbol vectors. This constraint imposes that, at the input of each of the $M_k$ RX RF chains, the self-interference signal cannot be larger than $\lambda_{\rm A}$. Notice also that in $\mathcal{OP}1$ we have not included a constraint similar to $({\rm C4})$ for simplification purposes, since removing this constraint allow us to solve $\mathcal{OP}1$ by decoupling the maximization of the uplink and downlink rates, while implicitly achieving some level of self-interference mitigation via the RX beamformer $\u_k$. The reason for the latter is that, since self interference degrades the uplink rate, an appropriate design for maximizing the instantaneous uplink rate would naturally result in self-interference reduction. Finally, we have included a constraint on the norm of $\u_k$ to avoid having solutions for $\u_k$ that result in undesired amplification of the received signals (from node $m$, self interference, and AWGN).

Since downlink communication is usually more rate demanding than the uplink, we propose to tackle $\mathcal{OP}1$ in the following decoupled way. First, we solve for $\C_k$ and $\V_k$ that maximize the instantaneous downlink rate subject to the relevant constraints for these unknown variables. More specifically, we formulate the following optimization subproblem for the design of $\C_k$ and $\V_k$:  
\begin{equation*}\label{eq:optim2}
\begin{split}
  \mathcal{OP}2: &\max_{\C_k,\V_k} \mathcal{R}_{\rm DL}\left(\V_k\right)~~\textrm{s.t.}~~({\rm C1}),\,\,({\rm C2}),
	\\& \hspace{0.4cm}\|[\widetilde{\H}_{k,k}\V_k]_{(i,:)}\|^2\leq\lambda_{\rm A} \hspace{0.15cm} \forall i=1,2,\ldots,M_k.
\end{split}
\end{equation*}
To solve the latter problem we adopt an alternating optimization approach. Specifically, supposing that the available number of analog canceller taps $N$ and a realization of $\C_k$ satisfying the constraint $({\rm C2})$ are given, we seek for $\V_k$ maximizing the downlink rate while meeting the constraint $({\rm C1})$ and the constraint for the residual self interference after analog cancellation. The latter procedure is repeated for all allowable realizations of $\C_k$ for the given $N$ in order to find the best pair of $\C_k$ and $\V_k$ solving $\mathcal{OP}2$. The solution for $\V_k$ given $\C_k$ is summarized in Algorithm~\ref{DL_Precoding}, where the TX digital beamformer is constructed as $\V_k=\F_k\G_k$ with $\F_k\in\Compl^{N_k\times\alpha}$ and $\G_k\in\Compl^{\alpha\times d_k'}$, where $\alpha$ is a positive integer taking the values $1\leq\alpha\leq\alpha_{\max}$ and $d_k'\leq\min\{M_q,\alpha\}$. In general, $\alpha_{\max}=N_k$, however, for large transmission powers and strictly small values for $\lambda_{\rm A}$ it is advisable to set $\alpha_{\max}=\min\{M_q,N_k\}$. In principle, there exists a trade off for $\alpha$: the larger its value is, the higher the downlink rate and self interference are. For each $\alpha$ value we adopt a similar approach to \cite{SofNull_2016} for the design of $\V_k$. Particularly, its $\F_k$ component aims at minimizing the impact of the residual self-interference MIMO channel $\widetilde{\H}_{k,k}$, whereas the goal of the $\G_k$ component is to maximize the rate of the effective downlink channel $\H_{q,k}\F_k$. For the cases where $\H_{q,k}\F_k$ is a MIMO channel, $\G_k$ is given by the open-loop or closed-loop precoding for this channel derived using \cite{J:Telatar_Waterfilling}, depending on whether $\H_{q,k}$ is unknown or known, respectively, at the transmit node $k$. As seen from the algorithmic steps included in Algorithm~\ref{DL_Precoding}, we search for the largest allowable value for $\alpha$ maximizing the downlink rate while meeting the self-interference constraint concerning the uplink communication.

For a given number of analog canceller taps $N$ there are in total $\binom{M_kN_k}{N}$ ways to connect them from the available $N_k$ TX antenna elements to the available $M_k$ RX antenna elements. Each of those ways refers to a different realization of the $\C_k$ matrix and corresponds to a specific placement of the $N$ tap values inside $\C_k$; its remaining $M_kN_k-N$ elements need to be set to zeros. One reasonable $\C_k$ realization intended for satisfying the self-interference constraint in $\mathcal{OP}2$ is to obtain $\L_1$, $\L_2$, and $\L_3$ such that the resulting analog canceller matrix $\C_k$ has the $N$ tap values at the same elements with the $N$ largest in amplitude elements of $\H_{k,k}$. This $\C_k$ realization will result in cancelling the largest self-interference components. For example, suppose that $N_k=3$, $M_k=4$, and $N=2$ and that $[\H_{k,k}]_{2,1}$ and $[\H_{k,k}]_{4,2}$ are the two largest in amplitude elements of $\H_{k,k}$. In this case, we may design $\L_2={\rm diag}\{[[\H_{k,k}]_{2,1} [\H_{k,k}]_{4,2}]\}$, $[\L_1]_{1,1}=[\L_1]_{2,2}=1$, and $[\L_3]_{2,1}=[\L_3]_{4,2}=1$. Other reasonable realizations for $\C_k$ include the orderly column-by-column and row-by-row placement of the available $N$ tap values starting with the columns and rows, respectively, of $\H_{k,k}$ having the largest Euclidean norms. For example, suppose that $N_k=3$, $M_k=4$, and $N=3$, then having the three tap values placed at the $i$-th row of $\C_k$ will focus on reducing the self interference received at the $i$-th RX antenna element.
\begin{algorithm}[t!]\caption{TX Digital Precoding}\label{DL_Precoding}
\begin{algorithmic}[1]
\Statex \textbf{Input:} ${\rm P}_k$, $\H_{k,k}$, and $\H_{q,k}$ as well as a realization of $\C_k$ for a given $N$ satisfying the constraint $({\rm C2})$.
\State Obtain $\D_k$ including the $N_k$ right-singular vectors 
       \Statex of $\widetilde{\H}_{k,k}\triangleq\H_{k,k}+\C_k$ corresponding to the singular  
	  	 \Statex values in descending order.
\For{$\alpha=\alpha_{\max},\alpha_{\max}-1,\ldots,2$}
		\State Set $\F_k=[\D_k]_{(:,N_k-\alpha+1:N_k)}$.
		\State Set $\G_k$ as the optimum precoding for the effective
		       \Statex \hspace{0.55cm}downlink MIMO channel $\H_{q,k}\F_k$ given ${\rm P}_k$.
		\If{$\|[\widetilde{\H}_{k,k}\F_k\G_k]_{(i,:)}\|^2\leq\lambda_{\rm A}$ $\forall i=1,2,\ldots,M_k$,}
			 \State Output $\V_k=\F_k\G_k$ and stop the algorithm.
		\EndIf
\EndFor
\State Set $\F_k=[\D_k]_{(:,N_k)}$ and $\G_k={\rm P}_k^{1/2}$.
\If{$|[\widetilde{\H}_{k,k}\F_k\G_k]_{i}|^2\leq\lambda_{\rm A}$ $\forall i=1,2,\ldots,M_k$,}
			 \State Output $\V_k=\F_k\G_k$ and stop the algorithm.
\Else
			 \State Output that the $\C_k$ realization does not meet 
			 \Statex \hspace{0.55cm}the residual self-interference constraint.
\EndIf
\end{algorithmic}
\end{algorithm}


Using $\C_k$ and $\V_k$ from the solution of $\mathcal{OP}2$ we now proceed to the design of $\u_k$ that maximizes the instantaneous uplink rate. In particular, we formulate the following optimization subproblem for the RX digital combiner:
\begin{equation*}\label{eq:optim3}
  \mathcal{OP}3: \max_{\u_k} \gamma_k\left(\C_k,\V_k,\u_k\right)~~\textrm{s.t.}~~\|\u_k\|^2=1.
\end{equation*}
The $\h_{k,m}$ and $\widetilde{\H}_{k,k}$ appearing in \eqref{Eq:SINR} and thus included in $\mathcal{OP}3$ are assumed to be available at node $k$ through appropriately designed training phases. With the availability of this channel knowledge, it can be shown that the $\u_k$ solving $\mathcal{OP}3$ is given using \cite{Sadek2007_TSP_all} by the eigenvector that corresponds to the maximum eigenvalue of the matrix $\A\in\Compl^{M_k\times M_k}$, which is defined as
\begin{equation}\label{Eq:u_k}
\A \triangleq {\rm P}_m\left(\widetilde{\H}_{k,k}\V_k\V_k^{\rm H}\widetilde{\H}_{k,k}^{\rm H}+\sigma_k^2\I_{M_k}\right)^{-1}\h_{k,m}\h_{k,m}^{\rm H}.
\end{equation}
We note that for the practical case of imperfect analog cancellation, significant gains with RX digital combining are feasible only when it holds $M_k-d_k\geq d_m$.

According to the presented solutions of $\mathcal{OP}2$ and $\mathcal{OP}3$ for the joint design of analog cancellation and digital beamforming, the resulting values for $\C_k$, $\V_k$, and $\u_k$ are functions of channel matrices. This implies that the update of the TX and RX digital beamforming settings as well as the settings of the analog canceller (values for the taps and MUX/DEMUX configurations) depends on the coherence time of the involved wireless channels.

\section{Simulation Results and Discussion}\label{sec:Results}
The performance of the wireless communication scenario in Fig.~\ref{fig:FD_MIMO} using the FD MIMO design presented in Sec$.$~\ref{sec:Solution} is investigated in this section. In Sec$.$~\ref{subsec:Schemes} that follows we describe the SotA solutions with which the proposed design will be compared. The simulations parameters and assumptions are detailed in Sec$.$~\ref{subsec:Parameters}, whereas the hardware complexity, self-interference mitigation capability, and achievable rate results are presented in Sec$.$~\ref{subsec:Complexity_Rates}.  

\subsection{Compared FD MIMO Designs}\label{subsec:Schemes}       
We compare our novel FD MIMO design versus the combined cancellation and spatial suppression design presented in~\cite{Rii11_all} and the digital beamforming design recently proposed in~\cite{SofNull_2016}. We note that the designs presented in \cite{Atz16_all, Zha12_all} were not considered in the results that follow due to the fact they are only applicable to uplink and downlink communication with $d_k=d_m=1$, whereas our proposed design holds for $d_k\geq1$. A detailed description of the FD MIMO designs that will be compared is provided below.


\textit{Design 1: Proposed with $N$ taps.} This is our proposed FD MIMO design for the case of $N$ taps for analog cancellation. Compared with the SotA analog canceller architectures in \cite{Bha14, Kol16_all} which require at least $M_kN_k$ taps, our analog canceller results in $100(1-N/(M_kN_k))$\% reduction in the required number of taps. The TX and RX digital beamformers as well as the settings for the analog canceller at node $k$ are computed as presented in Sec$.$~\ref{sec:Solution}; particularly for the computation of $\G_k$ we have adopted open-loop MIMO precoding when needed.

\textit{Design 2: SotA with $M_kN_k$ taps.} This refers to a combination of time domain cancellation with spatial suppression as proposed in \cite{Rii11_all}. The TX digital beamformer is designed to minimize the self interference from the transmit side by using null space projection \cite{Rii11_all}. The RX digital beamforming is a minimum mean squared error (MMSE) filter, hence we compute it in the same way we compute $\u_k$ as explained in Sec$.$~\ref{sec:Solution}. The time domain cancellation is an analog canceller which requires a total of $M_kN_k$ taps, i$.$e$.$, one tap per TX-RX RF chain as in the SotA schemes \cite{Bha14, Kol16_all}. We have made the same assumptions for the analog canceller taps for this design as in our proposed analog canceller. 

\textit{Design 3: SotA with $0$ taps.} This is the SoftNull method presented in \cite{SofNull_2016} that uses no analog cancellation relying solely on TX digital beamforming to reduce self interference at the RX antenna elements of node $k$. Any residual self interference is handled by the RX digital combiner. The combiner $\u_k$ used in the previous two designs is used for the latter purpose.



\subsection{Simulation Parameters}\label{subsec:Parameters}       
We simulate the ergodic rate performance for a FD MIMO system as in Fig.~\ref{fig:FD_MIMO} with $M_k=N_k=4$ and $M_q=\{1,4\}$. We have assumed Rayleigh fading and a path loss of $110$dB for both the downlink $\H_{q,k}$ and uplink $\h_{k,m}$ communication channels. The self-interference channel $\H_{k,k}$ is assumed to be subject to Ricean fading with $K$-factor equal to $35$dB and path loss of $40$dB \cite{Dua12_all}. All involved wireless channels are assumed to be independent and identically distributed, and perfectly estimated at the receivers. Both the downlink and uplink transmit powers ${\rm P}_k$ and ${\rm P}_m$ are set between $20$dBm and $40$dBm. The noise floor at node $q$ is $-90$dBm and at node $k$ is $-110$dBm; the latter values are typical ones for mobile terminals and small cell base stations. Following the findings of \cite{Sab14_all} we consider a $14$-bit ADC at node $k$ that renders digital self-interference mitigation of approximately $50$dB feasible. This means that for the noise floor of $-110$dBm at node $k$ the residual self interference after analog cancellation (i$.$e$.$, at each RX RF chain's input) must be less than $-60$dBm. For the analog canceller taps, we assume non-ideal tap hardware, where each tap is set with steps of $0.02$dB for attenuation and $0.13^{\rm o}$ for phase; these values match the step values reported in \cite{Kol16_all}. Thus, for each tap in our simulation, the phase setting has a random phase error uniformly distributed between $-0.065^{\rm o}$ and $0.065^{\rm o}$, and the amplitude setting has a random amplitude error uniformly distributed between $-0.01$dB and $0.01$dB. 


\subsection{Hardware Complexity, Self-Interference Mitigation Capability, and Achievable Rates}\label{subsec:Complexity_Rates}  
\begin{figure}
	\begin{center}
	\includegraphics[width=3.4in]{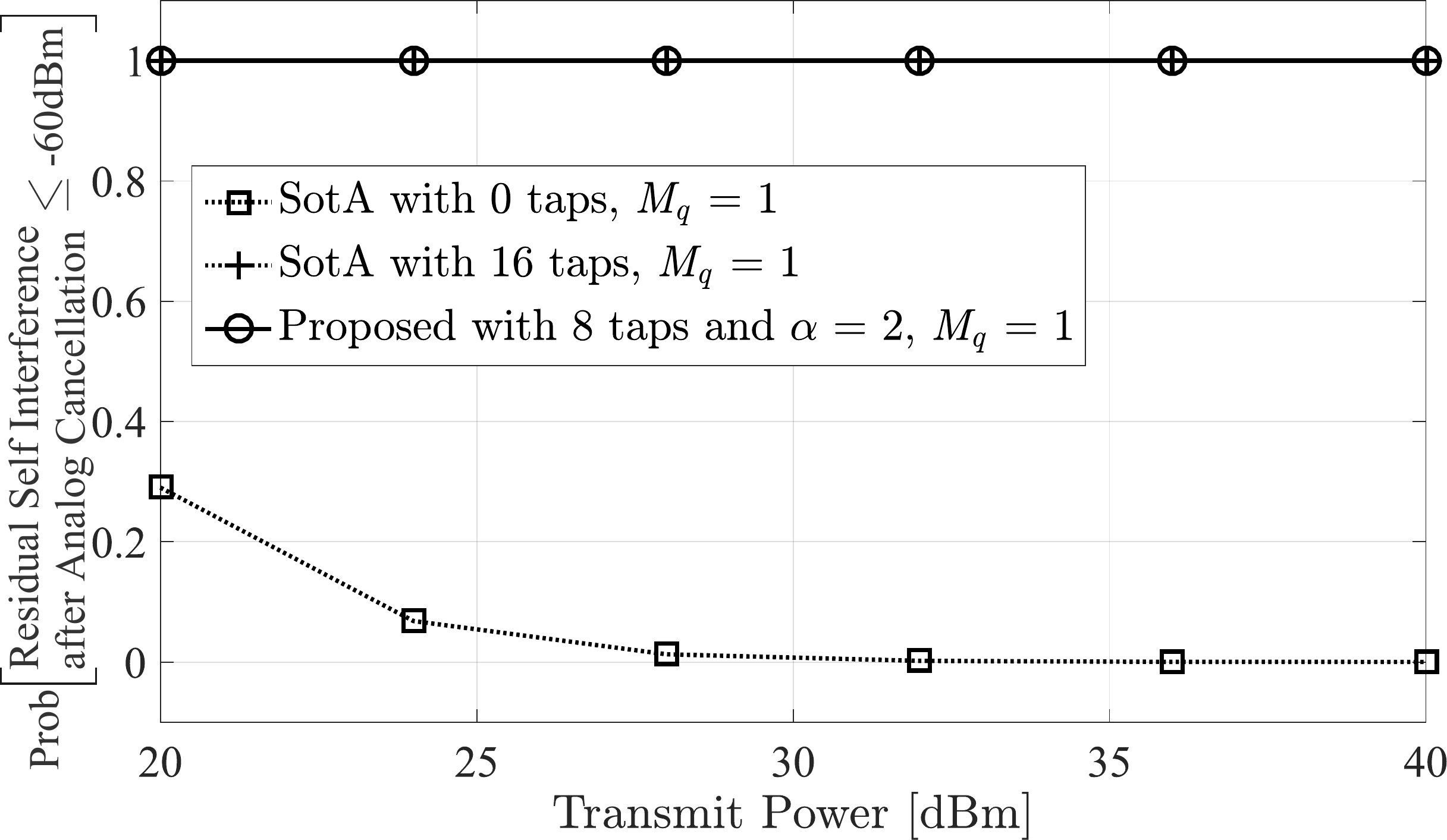}
	\caption{Probability of the residual self interference at the RX RF chains being less or equal to $\lambda_{\rm A}=-60$dBm versus the downlink transmit power for $M_k=N_k=4$ and $M_q=1$.}
	\label{fig:SI_Mitigation_Prob}
	\end{center}
\end{figure}
We have considered a \textit{Proposed with 8 taps} design and ran Algorithm~\ref{DL_Precoding} for the parameters in Sec$.$~\ref{subsec:Parameters}, $N=8$ taps, and different $\C_k$ realizations. We found that setting $\alpha=2$ when designing the TX digital beamformer and placing the $8$ taps in the first two rows of $\C_k$ results in the maximum downlink rate, while meeting the threshold $\lambda_{\rm A}=-60$dBm. The latter is showcased in Fig$.$~\ref{fig:SI_Mitigation_Prob}, where we set $M_q=1$ and depict the probability of achieving residual self interference less or equal to $\lambda_{\rm A}$ as a function of the considered transmit powers. Apart from the proposed TX beamforming design, we also sketch within Fig$.$~\ref{fig:SI_Mitigation_Prob} the probability curves with the precoders of \textit{SotA with 16 taps} and \textit{SotA with 0 taps}; for the latter design we have used $\alpha=2$ as in the proposed one. As shown, \textit{SotA with 0 taps} does not provide enough self-interference mitigation, and hence, it should not be used for the considered transmit power range. The proposed design, however, guarantees meeting $\lambda_{\rm A}$ with $50$\% less taps than the \textit{SotA with 16 taps} design does.
\begin{figure}
	\begin{center}
	\includegraphics[width=3.4in]{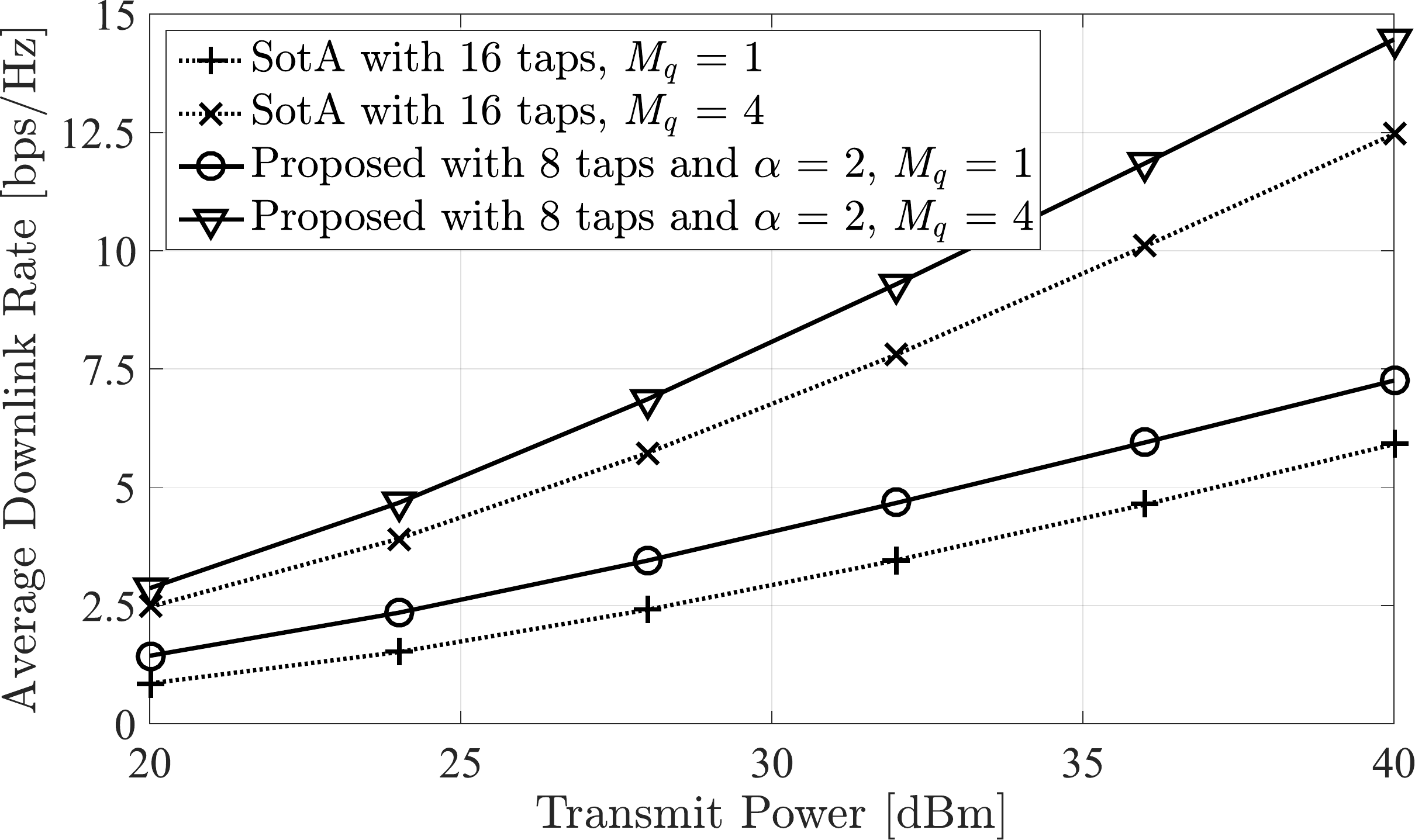}
	\caption{Average downlink rate as a function of the downlink transmit power for $M_k=N_k=4$ and $M_q=\{1,4\}$.}
	\label{fig:DL_Rates}
	\end{center}
\end{figure}

\begin{figure}
	\begin{center}
	\includegraphics[width=3.4in]{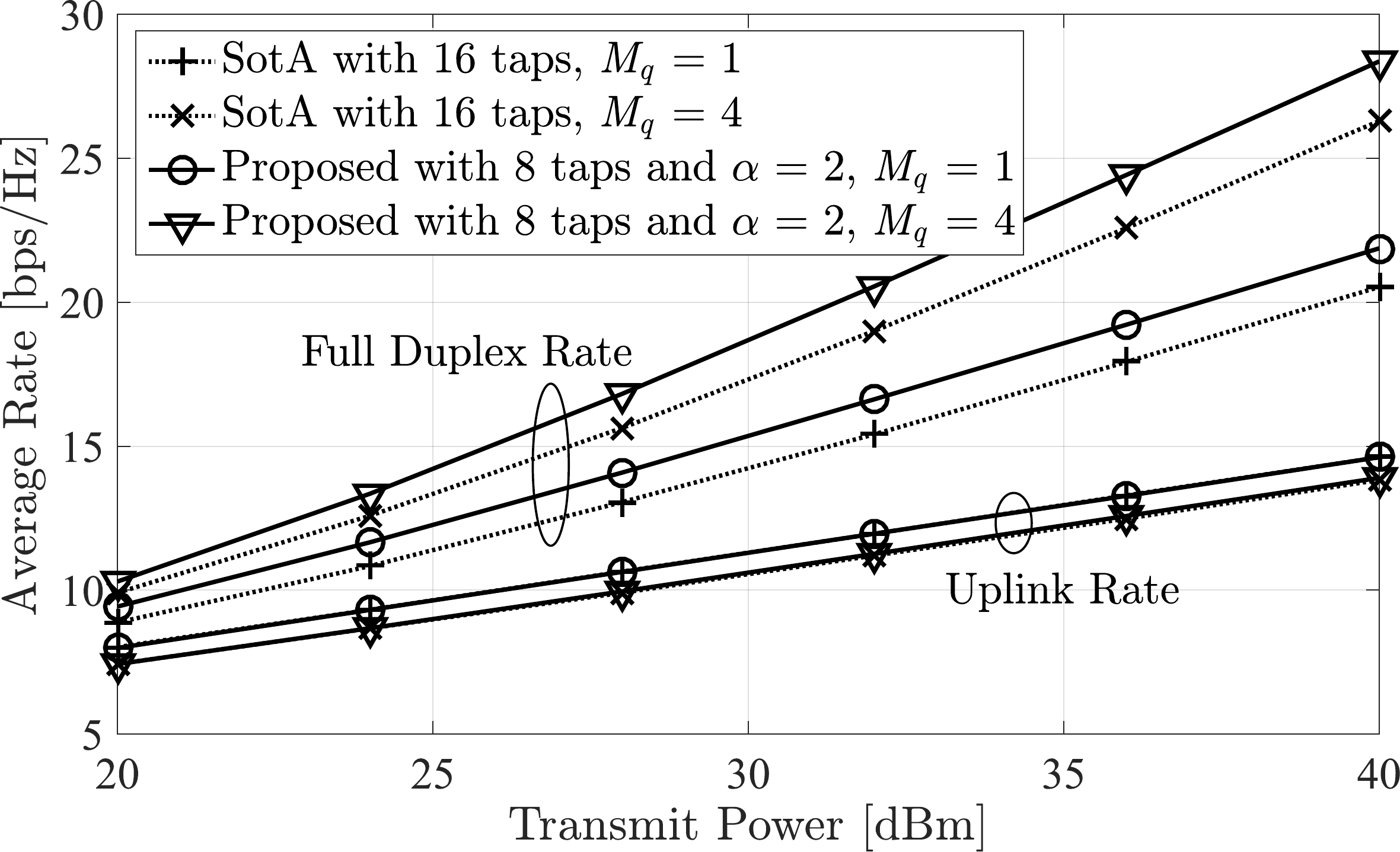}
	\caption{Average uplink and FD rates as functions of the transmit power for $M_k=N_k=4$ and $M_q=\{1,4\}$.}
	\label{fig:UL_FD_Rates}
	\end{center}
\end{figure}

The ergodic downlink, uplink, and FD rates in bps/Hz for both the designs \textit{SotA with 16 taps} and \textit{Proposed with 8 taps} and $\alpha=2$ are plotted in Figs$.$~\ref{fig:DL_Rates} and \ref{fig:UL_FD_Rates} as functions of the considered transmit powers. It is shown in Fig$.$~\ref{fig:DL_Rates} that the proposed TX beamformer delivers higher downlink rates than the SotA one, and as expected, this performance superiority increases with increasing ${\rm P}_k$ and/or $M_q$. Both designs achieve the same uplink rate irrespective of $M_q$ as illustrated in Fig$.$~\ref{fig:UL_FD_Rates}. As expected, increasing $M_q$ increases self interference, thus degrades the uplink rate. Interestingly, this uplink performance degradation is kept very low with the proposed design. We thus conclude that, compared to SotA, the proposed design can achieve higher rates with $50$\% less taps.


\section{Conclusion and Future Work}\label{sec:Concl}
In this paper, we have presented a novel self-interference mitigation scheme for FD MIMO systems with reduced hardware complexity. The proposed scheme includes a novel multi-tap analog canceller architecture whose configuration is jointly designed with digital beamforming. The main simplification of the analog canceller hardware was obtained via the use of multiplexers for signal routing among the TX and RX RF chains and the reduced number of taps, and the co-design of the tap values and configuration of the multiplexers with the TX and RX beamforming filters. We have presented a general optimization framework for the latter joint design and detailed a specific solution targeting at the FD rate maximization. The performance evaluation results demonstrated that our proposed design can be implemented with less taps than SotA ones while achieving larger FD rates. For future work, we intend to extend the proposed design to wideband channels and apply the proposed framework to analog cancellation solutions based on auxiliary TXs as in \cite{Dua14_all}.


\bibliographystyle{IEEEtran}
\bibliography{IEEEabrv,refs_FD_architecture}

\end{document}